\documentstyle[twoside,fleqn,espcrc2]{article}

\newcommand{\AmS}{{\protect\the\textfont2
  A\kern-.1667em\lower.5ex\hbox{M}\kern-.125emS}}

\begin{document}
\begin{titlepage}
\thispagestyle{empty}

\phantom{anchor}
\vspace*{-50mm}

\begin{flushright} \large
 {\tt University of Bergen, Department of Physics}    \\[2mm]
 {\tt Scientific/Technical Report No.1996-06}    \\[2mm]
 {\tt ISSN 0803-2696} \\[5mm]
 {hep-ph/9605453} \\[5mm]
 {May 1996}           \\
\end{flushright}

\vspace*{2cm}

\begin{center}
{\bf \Large
One-loop results for three-gluon vertex in arbitrary gauge 
and dimension${}^*$}
\end{center}
\vspace{1cm}
\begin{center}
\large
A.~I.~Davydychev$^{a}$, \ \
P.~Osland$^{a}$
 \ \ and \ \
O.~V.~Tarasov$^{b}$

\vspace{1cm}
$^{a}${\em
       Department of Physics, University of Bergen, \\
       All\'{e}gaten 55, N-5007 Bergen, Norway}
\\
\vspace{.3cm}
$^{b}${\em
       IfH, DESY-Zeuthen,
       Platanenallee 6, D-15738 Zeuthen,
       Germany}
\\
\vspace{.3cm}

\vspace{30mm}

\begin{minipage}{120mm}
\large
We review the calculation of one-loop contributions
to the three-gluon vertex, for arbitrary (off-shell) external momenta,
in arbitrary covariant gauge and in arbitrary space-time dimension.
We discuss how one can get the results for all on-shell limits
of interest directly from the general off-shell expression.
\end{minipage}

\end{center}
\vspace{30mm}

\begin{minipage}{14cm}
\begin{minipage}{60mm}
\hrulefill
\end{minipage}

${}^*$ Presented at 
{\sl QCD and QED in Higher Order}, 1996 Zeuthen Workshop on 
Elementary Particle Theory, Rheinsberg, April 21--26, 1996,
to appear in Nucl.\ Phys.\ B (Proc. Suppl.)
\end{minipage}

\end{titlepage}
\newpage

\title{One-loop results for three-gluon vertex in arbitrary
       gauge and dimension}

\author{A.~I.~Davydychev${}^{\rm a}$\thanks{Permanent address:
                  Institute for Nuclear Physics,
                  Moscow State University,
                  119899, Moscow, Russia.} , 
        P.~Osland\address{Department of Physics, University of Bergen,
        All\'{e}gaten 55, N-5007 Bergen, Norway}
        and
        O.~V.~Tarasov\address{IfH, DESY-Zeuthen, Platanenallee 6, 
                  D-15738 Zeuthen, Germany.}\thanks{On leave from
                  Joint Institute for Nuclear Research,
                  141980, Dubna, Russia.}}


\begin{abstract}
We review the calculation of one-loop contributions 
to the three-gluon vertex, for arbitrary (off-shell) external momenta,
in arbitrary covariant gauge and in arbitrary space-time dimension.
We discuss how one can get the results for all on-shell limits 
of interest directly from the general off-shell expression.
\end{abstract}

\maketitle

\section{INTRODUCTION}

The three-gluon vertex is the basic object of Quantum Chromodynamics
\cite{QCD} ``responsible'' for its non-Abelian nature.
In the standard covariant-gauge perturbation theory, the one-loop 
contributions to 
the three-gluon vertex have been studied in a number of papers. 
Celmaster and Gonsalves \cite{CG} presented
the one-loop result for the vertex, for off-shell gluons,
restricted to the symmetric case, $p_1^2=p_2^2=p_3^3$,
in an arbitrary covariant gauge.
Then, Ball and Chiu \cite{BC2} considered the general off-shell case,
but restricted to the Feynman gauge.
Later, various on-shell results have also been given,
by Brandt and Frenkel \cite{BF}, 
restricted to the infrared-singular parts only 
(in an arbitrary covariant gauge), 
and by Nowak, Prasza{\l}owicz and S{\l}omi{\'n}ski 
\cite{NPS}, who also gave the finite parts for the case of
two gluons being on-shell (in the Feynman gauge). 

In a recent paper \cite{DOT}, we have completed the investigation
of the one-loop three-gluon vertex in covariant gauge
(for the case when massless quarks are considered).
We have obtained results which are valid 
for {\em arbitrary} values of the space-time dimension
and the gauge parameter.
Apart from the three-gluon vertex itself, we have also considered 
the ghost-gluon vertex (as well as two-point functions), 
to be able to check
that all these quantities obey the Ward--Slavnov--Taylor
identity for the three-gluon vertex.
The present paper contains a brief overview of those results.

\section{NOTATION}

The lowest-order three-gluon vertex is well known,
\begin{eqnarray}
\label{2eq:tree-level}
-\!\mbox{i} g f^{a_1 a_2 a_3} 
\bigl[     g_{\mu_1 \mu_2} (p_1 \!-\! p_2)_{\mu_3}
         \!+\! g_{\mu_2 \mu_3} (p_2 \!-\! p_3)_{\mu_1} 
\hspace*{-2mm} 
\nonumber \\
         + g_{\mu_3 \mu_1} (p_3 - p_1)_{\mu_2} \bigr] , 
\end{eqnarray}
where $p_1, p_2$ and $p_3$ are the momenta of the gluons,
all of which are ingoing, $p_1+p_2+p_3=0$.
The $f^{a_1 a_2 a_3}$ 
are the totally antisymmetric colour structures corresponding
to the adjoint representation of the gauge group.
They can
be extracted from the general three-gluon vertex by defining
\begin{eqnarray}
\label{ggg}
\Gamma_{\mu_1 \mu_2 \mu_3}^{a_1 a_2 a_3}(p_1, p_2, p_3) 
\hspace{27mm}
\nonumber \\
\equiv  - \mbox{i} \; g \;
f^{a_1 a_2 a_3} \; \Gamma_{\mu_1 \mu_2 \mu_3}(p_1, p_2, p_3) .
\end{eqnarray}
Since the colour structures $f^{a_1 a_2 a_3}$ are antisymmetric, 
$\Gamma_{\mu_1 \mu_2 \mu_3}(p_1, p_2, p_3)$ must also be {\em antisymmetric}
under any interchange of a pair of gluon momenta and the corresponding
Lorentz indices.

When one calculates one-loop (and higher) contributions 
to the three-gluon vertex,
other tensor structures
arise, in addition to the lowest-order expression 
(\ref{2eq:tree-level}), and the general tensor decomposition 
should be considered. 
We use the symmetric decomposition of the general 
three-gluon vertex, proposed 
in \cite{BC2},
\begin{eqnarray}
\label{BC-ggg}
&&\hspace*{-9mm} \Gamma_{\mu_1 \mu_2 \mu_3}(p_1, p_2, p_3) 
\nonumber \\
&\hspace*{-6mm}=& \hspace{-6mm} 
A(p_1^2, p_2^2; p_3^2)\; g_{\mu_1 \mu_2} (p_1-p_2)_{\mu_3}
\nonumber \\ 
&\hspace*{-6mm}+& \hspace{-6mm} 
B(p_1^2, p_2^2; p_3^2)\; g_{\mu_1 \mu_2} (p_1+p_2)_{\mu_3}
\nonumber \\
&\hspace*{-6mm}-& \hspace{-6mm}  
C(p_1^2, p_2^2; p_3^2)
\left( (p_1 p_2) g_{\mu_1 \mu_2}\!\!-\!{p_1}_{\mu_2} {p_2}_{\mu_1} \right)
\!(p_1\!-\!p_2)_{\mu_3}
\hspace*{-1mm}
\nonumber \\
&\hspace*{-6mm}+& \hspace{-6mm}  
\textstyle{1\over3} S(p_1^2, p_2^2, p_3^2)
\left( {p_1}_{\mu_3} {p_2}_{\mu_1} {p_3}_{\mu_2}
        \!+\! {p_1}_{\mu_2} {p_2}_{\mu_3} {p_3}_{\mu_1} \right)
\nonumber \\
&\hspace*{-6mm}+& \hspace{-6mm} 
F(p_1^2, p_2^2; p_3^2)
\left( (p_1 p_2) g_{\mu_1 \mu_2} - {p_1}_{\mu_2} {p_2}_{\mu_1} \right)
\nonumber \\
&& \hspace*{21mm}
\times
\left( {p_1}_{\mu_3} (p_2 p_3) - {p_2}_{\mu_3} (p_1 p_3) \right)
\nonumber \\
&\hspace*{-6mm}+& \hspace{-6mm} 
H(p_1^2, p_2^2, p_3^2)
\left[ -g_{\mu_1 \mu_2}
\left( {p_1}_{\mu_3} (p_2 p_3) \!-\! {p_2}_{\mu_3} (p_1 p_3) \right)
\right.
\hspace*{-1mm}
\nonumber \\
&& \hspace*{10mm}
+ \textstyle{1\over3}
\left.
\left( {p_1}_{\mu_3} {p_2}_{\mu_1} {p_3}_{\mu_2}
\!-\! {p_1}_{\mu_2} {p_2}_{\mu_3} {p_3}_{\mu_1} \right) \right]
\nonumber \\
&+& \left\{ \mbox{cyclic permutations}\right\}  .
\end{eqnarray}
The $A$, $C$ and $F$ functions are symmetric in the first two
arguments, the $H$ function is totally symmetric, $B$ 
is antisymmetric
in the first two arguments, while $S$ is antisymmetric
with respect to interchange of any pair of arguments.
The $F$ and $H$ contributions 
are totally transverse, i.e.\ they give zero when contracted with
any of ${p_1}_{\mu_1}$, ${p_2}_{\mu_2}$ or ${p_3}_{\mu_3}$.
At the zero-loop level, the only non-vanishing function is 
$A^{(0)}=1$ (cf. eq.~(\ref{2eq:tree-level}) ).

The set of one-loop diagrams contributing to the three-gluon vertex
contains the gluon, ghost and quark ``triangles'', as well as
three ``bubbles'' involving four-gluon vertices.
The tensor structure of the gluon propagator (with the momentum $p$) is
$( g_{\mu_1 \mu_2} - \xi p_{\mu_1} p_{\mu_2}/p^2 )$,
where $\xi$ is the co\-va\-ri\-ant-gauge parameter (taking $\xi=0$
corresponds to the Feynman gauge). In what follows, we shall
use the superscript ``$(1,\xi)$'' for the contributions in arbitrary gauge
(without quark loops), ``$(1,0)$'' for the same contributions in the
Feynman gauge, and ``$(1,q)$'' for the contribution of the quark loops.
The superscript ``$(1)$'' refers to the sum of all one-loop contributions.  

In the one-loop calculation,
the following scalar integrals (corresponding to the 
triangle diagrams) are involved:
\[
J(\nu_1, \nu_2, \nu_3) \equiv \!\int\!
 \frac{\mbox{d}^n q}{ ((p_2 -q )^2)^{\nu_1}  ((p_1 +q )^2)^{\nu_2}
      (q^2)^{\nu_3} }
\]
where $n=4-2\varepsilon$ is the space-time dimension.
All such integrals occurring in the calculation
of the one-loop three-gluon vertex  
can be algebraically reduced to one non-trivial integral,
\begin{equation}
\label{J(1,1,1)}
J(1,1,1) = {\mbox{i}} \pi^{n/2} \;
\eta \;
\varphi(p_1^2,p_2^2,p_3^2) ,
\end{equation}
where $\varphi(p_1^2,p_2^2,p_3^2) \equiv \varphi$ is a totally
symmetric function,
and three two-point integrals, $J(0,1,1)$, $J(1,0,1)$ and $J(1,1,0)$, 
which can be expressed in terms of a power-like function
\begin{equation}
\label{kappa}
\kappa(p_i^2) \equiv \kappa_i
= - \frac{2}{(n-3) (n-4)} \; (-p_i^2)^{(n-4)/2}
\end{equation}
as, e.g.,
$J(1,1,0) = {\mbox{i}} \pi^{n/2} \; \eta \; \kappa(p_3^2)$,
and similarly for $J(0,1,1)$ and $J(1,0,1)$.
In these equations,
$\eta$ denotes a factor constructed of $\Gamma$ functions,
\begin{equation}
\label{eta}
\eta \equiv 
\Gamma^2({\textstyle{n\over2}}-1) \Gamma(3-{\textstyle{n\over2}})/
\Gamma(n-3) .
\end{equation}
Note that integrals $J(\nu_1,\nu_2,\nu_3)$ with two non-positive
powers $\nu_i$ vanish in dimensional regularization \cite{dimreg}.
For $\xi\neq 0$, we also get integrals with some of the powers
of the denominators equal to two.
However, with the help of the integration-by-parts technique \cite{ibp}
these integrals can be algebraically reduced to the above basic set
(see ref.~\cite{JPA}). 

For the integral (\ref{J(1,1,1)}), a one-dimensional
integral representation \cite{UD3}
(which is valid for arbitrary $\varepsilon$) can be derived,
\[
-\frac{\mbox{i} \pi^{2-\varepsilon} \eta}{(-p_3^2)^{1+\varepsilon}} \;
\frac{1}{\varepsilon} \int\limits_0^1
\frac{\mbox{d}\sigma\;\sigma^{-\varepsilon}
                 \left( (y\sigma)^{-\varepsilon}
                \!-\! (x/\sigma)^{-\varepsilon} \right)}
      {\left( y \sigma^2 \!+\! (1\!-\!x\!-\!y)\sigma 
                         \!+\! x \right)^{1-\varepsilon}} .
\]
Using the connection between massive and massless diagrams \cite{DT2}
and the known expressions for two-loop massive vacuum diagrams
\cite{walnut}, the result for $J(1,1,1)$ (in arbitrary space-time dimension)
can be presented in terms of $_2F_1$ hypergeometric functions.
In four dimensions, the off-shell result can be expressed in terms
of dilogarithms (see e.g. in \cite{'tHV-79,BC2,JPA}).

\section{OFF-SHELL RESULTS}

When calculating the diagrams, we used the standard technique
of tensor decomposition \cite{tensor}, reducing the result to
combinations of scalar integrals multiplying the tensor
structures constructed from the external momenta.
Then, we constructed the scalar functions (\ref{BC-ggg}) as certain
linear combinations of the coefficients of these tensor structures.
Finally, all scalar integrals were expressed in terms of the
above-mentioned basis of the four integrals.
While performing the calculations,
the {\sf REDUCE} system \cite{reduce} was heavily employed.

Before presenting selected results, we define two totally symmetric
combinations of the invariants formed from the external momenta,
\begin{eqnarray*}
{\cal{Q}} 
\equiv (p_1 p_2) \!+\! (p_1 p_3) \!+\! (p_2 p_3)
          = - \textstyle{1\over2} (p_1^2 \!+\! p_2^2 \!+\! p_3^2) , 
\hspace*{-5mm}
\\
{\cal{K}} 
\equiv p_1^2 p_2^2 - (p_1 p_2)^2 =
- \textstyle{1\over4} \lambda\left(p_1^2,p_2^2,p_3^2 \right) ,
\hspace{7mm}
\end{eqnarray*}
where $\lambda$ is the K\"allen function.

Let us first consider the one-loop contributions 
to the three-gluon vertex
(\ref{BC-ggg}) in the Feynman gauge ($\xi=0$), without the quark loops.
We shall use the standard notation $C_A$ for the Casimir constant,
$ f^{acd}f^{bcd} = C_A \, \delta^{ab}$, with $(C_A = N$ for the SU($N$) 
group), 
whereas the factor $\eta$ occurring in the results is defined by
eq.~(\ref{eta}).
 
The one-loop results for the functions
(\ref{BC-ggg}), for arbitrary value of the
space-time dimension $n$, are  
\begin{eqnarray}
\label{A(1,0)}
A^{(1,0)}(p_1^2, p_2^2; p_3^2) 
=\frac{g^2 \; \eta}{(4\pi)^{n/2}} \; \frac{C_A}{4 (n-1) \cal{K}}
\hspace{9mm} 
\nonumber \\
\times\! \bigl\{ (n-1) \left( p_3^2 + 3 (p_1 p_2)\right)
\hspace{29.5mm} 
\nonumber \\
\times\!
\left[ p_3^2 (p_1 p_2) \varphi \!+\! (p_1 p_3) \kappa_1
      \!+\! (p_2 p_3) \kappa_2 \!+\! p_3^2 \kappa_3 \right]
\hspace*{-2mm}
\nonumber \\
+ 4 (n-1) {\cal{K}} \left[ (p_1 p_2) \varphi + \kappa_3 \right]
\hspace{25mm}
\nonumber \\
- (3n-2) {\cal{K}} \left[ \kappa_1 + \kappa_2 \right]
\bigr\} , 
\hspace{30mm}
\\
\label{B(1,0)}
B^{(1,0)}(p_1^2, p_2^2; p_3^2) 
= - \frac{g^2 \; \eta}{(4\pi)^{n/2}} \;  
 \frac{C_A \; (p_1^2 - p_2^2)}{4 (n-1) \cal{K}} 
\hspace{2mm}
\nonumber \\
\times\!
\bigl\{ (n\!-\!1)
\left[ (p_1 p_3)(p_2 p_3) \varphi 
\right.
\hspace{30mm}
\nonumber \\
\left.
+ (p_1 p_3) \kappa_1 
       + (p_2 p_3) \kappa_2 
+ p_3^2 \kappa_3 \right]
\nonumber \\
+ (4n-3) {\cal{K}} \; 
(\kappa_1-\kappa_2)/(p_1^2 \!-\! p_2^2)
\bigr\} , 
\hspace{14mm}
\\
\label{C(1,0)}
C^{(1,0)}(p_1^2, p_2^2; p_3^2) 
= \frac{g^2 \; \eta}{(4\pi)^{n/2}} \;
 \frac{C_A}{4 (n-1) \cal{K}} 
\hspace{9mm}
\nonumber \\
\times\!
\bigl\{ \! 3 (n\!-\!1) \!
\left[  p_3^2 (p_1 p_2) \varphi \!+\! (p_1 p_3) \kappa_1
      \!+\! (p_2 p_3) \kappa_2 \!+\! p_3^2 \kappa_3 \! \right]
\hspace*{-8mm}
\nonumber \\
- 2\; (4n\!-\!3)\; {\cal{K}} \; 
(\kappa_1\!-\!\kappa_2)/(p_1^2 \!-\! p_2^2)
\bigr\} , 
\hspace{11mm}
\\
\label{S(1,0)}
S^{(1,0)}(p_1^2, p_2^2, p_3^2) = 0 , 
\hspace{35.5mm}
\\
\label{F(1,0)}
F^{(1,0)}(p_1^2, p_2^2; p_3^2) 
=  \frac{g^2 \; \eta}{(4\pi)^{n/2}} \; 
\frac{C_A}{4 (n-1) {\cal{K}}^3}
\hspace{7.5mm}
\nonumber \\
\times \! \bigl\{ 
2  \left[(n^2-1) (p_1 p_2) (p_1 p_3) (p_2 p_3)
\right.
\hspace{18mm}
\nonumber \\
\left.
+2 (n-2) p_3^2 {\cal{K}}-(n-7) (p_1 p_2) {\cal{K}}\right]
\nonumber \\
\times\! \left[ p_3^2 (p_1 p_2) \varphi \!+\! (p_1 p_3) \kappa_1
             \!+\! (p_2 p_3) \kappa_2 \!+\! p_3^2 \kappa_3 \right]
\hspace*{-2mm}
\nonumber \\
+2 {\cal{K}} \left[(n\!+\!1) (n\!-\!4) (p_1 p_3) (p_2 p_3)
                          \!-\!(5n\!-\!11) 
{\cal{K}}\right] 
\nonumber \\
\times\!
\left[(p_1 p_2) \varphi+\kappa_3\right]
\nonumber \\
       +2 p_3^2 {\cal{K}} \left[(n+1) (p_1 p_3) (p_2 p_3)
                 +(n-3) {\cal{K}}\right] \varphi
\hspace{4mm}
\nonumber \\
       +(4 n-7) {\cal{K}}^2  \left[ \kappa_1+\kappa_2 \right]
\hspace{33.5mm}
\nonumber \\
       +  {\cal{K}} \left[
       2 (n\!+\!1) (p_1 p_2) (p_1^2 \!-\! p_2^2)^2
       \!+\! (4n\!-\!3)  {\cal{K}}
(p_1\!\!-\!p_2)^2
\right]
\hspace*{-7.6mm} 
\nonumber \\
\times
(\kappa_1 \!-\! \kappa_2)/(p_1^2 \!-\! p_2^2) 
\bigr\} , \\
\label{H(1,0)}
H^{(1,0)}(p_1^2, p_2^2, p_3^2) 
= \frac{g^2 \; \eta}{(4\pi)^{n/2}} \; 
\frac{C_A}{2 (n-1) {\cal{K}}^3}
\hspace{7mm}
\nonumber \\
\times\!
\bigl\{ (n^2-1) (p_1 p_2)(p_1 p_3)(p_2 p_3)
\hspace{23mm}
\nonumber \\
\times
\left[ (p_1 p_2)(p_1 p_3)(p_2 p_3) \varphi + (p_1 p_2)(p_1 p_3) \kappa_1
\right. \nonumber \\
\left.
+ (p_1 p_2)(p_2 p_3) \kappa_2 + (p_1 p_3)(p_2 p_3) \kappa_3 \right]
\nonumber \\
- 3 (n\!-\!1) (p_1 p_2)(p_1 p_3)(p_2 p_3) {\cal{K}}
\left[ {\cal{Q}} \varphi
        \!+\! \kappa_1 \!+\!  \kappa_2 \!+\! \kappa_3 \right]
\hspace*{-6.5mm}
\nonumber \\
+ 2 (n-1) {\cal{K}}^3 \; \varphi
\hspace{44mm}
\nonumber \\
+ (n\!-\!2) {\cal{K}} \left[
p_1^2 \left( p_1^2 (p_2 p_3) \!+\! (p_1 p_2)(p_1 p_3) \right) \kappa_1
\right.
\hspace{2.5mm}
\nonumber \\
+ p_2^2 \left( p_2^2 (p_1 p_3) \!+\! (p_1 p_2)(p_2 p_3) \right) \kappa_2
\hspace{3.5mm}
\nonumber \\
\left.
+ p_3^2 \left( p_3^2 (p_1 p_2) \!+\! (p_1 p_3)(p_2 p_3) \right) \kappa_3
\right] \bigr\} .
\end{eqnarray}
When expanded around $n=4$, these results correspond to those
presented in \cite{BC2} (see in \cite{DOT} for the details
of this comparison).

In \cite{DOT}, the results for (massless) quark-loop contributions
are also presented. For example,
\[
A^{(1,q)}(p_1^2, p_2^2; p_3^2)
= \frac{g^2 \; \eta}{(4\pi)^{n/2}}  N_f T_R 
\frac{n-2}{n-1}  \left[ \kappa_1 + \kappa_2 \right] ,
\]
where $N_f$ is the number of quarks and 
$T_R={\textstyle{1\over8}}{\mbox{Tr}}(I)={\textstyle{1\over2}}$.
The result $S=0$ is valid also for the quark loop contributions,
even when the quarks are massive (moreover, it is valid in an arbitrary 
gauge).
It should be noted that presenting the results in arbitrary
dimension does not spoil their compactness, as compared with
the formulae expanded around $n=4$.

In an arbitrary gauge, the results for the scalar functions of the 
three-gluon vertex (\ref{BC-ggg}) are obviously less compact
than those in the Feynman gauge. We list one of them below,
for arbitrary value of the space-time dimension: 
\begin{eqnarray}
\label{A(1,xi)}
A^{(1, \xi)}(p_1^2, p_2^2; p_3^2)
=\frac{g^2 \; \eta}{(4\pi)^{n/2}} \; C_A \; 
\frac{1}{32 {\cal{K}}^2 \; p_1^2 \; p_2^2}
\hspace{2mm}
\nonumber \\
   \times\!\biggl\{ \!
\bigl[ p_1^2 p_2^2 {\cal{K}} \bigl(
        \left( 8-4 \xi-(n-2) (n-3) \xi^2 \right) p_3^2
\hspace{5mm}
\nonumber \\
       +2 \left( 12+4 (n-3) \xi+(n-3) \xi^2 \right) (p_1 p_2) \bigr)
\hspace*{-7mm}
\nonumber \\
\!+ \xi \left( (n\!-\!4) \xi\!+\!4 \right) {\cal{K}} {\cal{Q}}
\left( (n\!-\!3) (p_1 p_2) {\cal{Q}} \!-\! (n\!-\!4) {\cal{K}} \right)
\hspace*{-7mm}
\nonumber \\
+ \xi \left( (n\!-\!3) \xi + 2 \right)
(n\!-\!1) p_1^2 p_2^2 p_3^2 (p_1 p_2) {\cal{Q}}
\bigr]
\hspace{4mm}
\nonumber \\
\times\! \left[ p_3^2 (p_1 p_2) \varphi +(p_1 p_3) \kappa_1 
           +(p_2 p_3) \kappa_2 +p_3^2 \kappa_3 \right]
\hspace*{-6mm}
\nonumber \\
 -{\cal{K}} 
\bigl[ \left( (n-4) \xi+4 \right) {\cal{K}}
\bigl( \left( (n-4) \xi-8 \right) p_1^2 p_2^2
\hspace{5.5mm}
\nonumber \\
+\xi {\cal{Q}} \left( (n\!-\!2) p_3^2-2 (n\!-\!3) (p_1 p_2) \right)
\bigr)
\hspace*{-7mm}
\nonumber \\
-\xi \left( (n\!-\!3) \xi \!+\! 2 \right) 
(n\!-\!2) p_1^2 p_2^2 p_3^2 {\cal{Q}}
\bigr]
\left[ (p_1 p_2) \varphi \!+\! \kappa_3 \right]
\hspace*{-8mm}
\nonumber \\
+{\cal{K}}\; \varphi 
\bigl[ \xi \left( (n-4) \xi+4 \right)
{\cal{K}} 
\hspace{30mm}
\nonumber \\
\times\!
\bigl( (2n\!-\!7) p_1^2 p_2^2 p_3^2
+{\cal{Q}} \left( p_1^2 (p_1 p_3) \!+\! p_2^2 (p_2 p_3) \right)
\bigr)
\hspace*{-7mm}
\nonumber \\
+\xi \left( (n\!-\!3) \xi+2 \right) p_1^2 p_2^2 p_3^2
   \left( p_3^2 {\cal{Q}}-2 (n\!-\!4) {\cal{K}} \right) \bigr] 
\hspace*{-7mm}
\nonumber \\
-\frac{{\cal{K}}}{n\!-\!1} \bigl[ p_1^2 p_2^2 {\cal{K}}
 \bigl( 8 (3 n\!-\!2)
+4 (n\!-\!1) (5 n\!-\!17) \xi
\hspace*{-0.5mm}
\nonumber \\
-3 (n-1) (n-4) \xi^2 \bigr)
\nonumber \\
+\xi \left( (n-4) \xi+4 \right) (n-1) {\cal{K}} {\cal{Q}}^2
\hspace{18mm}
\nonumber \\
+\xi \left( (n\!-\!3) \xi \!+\! 2 \right) 
(n\!-\!1) p_1^2 p_2^2 p_3^2 {\cal{Q}} \bigr]
\left[ \kappa_1 \!+\! \kappa_2 \right] \biggr\} ,
\end{eqnarray}

One of the main technical problems we met in this calculation
was how to bring the results for arbitrary $\xi$ to a reasonably
short form, 
i.e.\ how to organize the result and which bases to choose. 
First, it was possible to get better factorization of the coefficients
by considering not $\varphi$ and $\kappa_i$ themselves but certain linear
combinations. 
Then, the next idea was
to try to use in some cases not only $p_1^2, p_2^2, p_3^2$ but 
also the scalar products $(p_1 p_2), (p_1 p_3)$
and $(p_2 p_3)$, together with the notation 
${\cal Q}$ and ${\cal K}$ for symmetric combinations. 
These tricks (as well as 
looking for proper combinations of $\xi$ and $n$) allowed us to
write the expressions in a much shorter form. 

There are some special values of the gauge parameter $\xi$
we would like to point out.
First of all, we see that the terms containing
$p_1^2, p_2^2$ or $p_3^2$ in the denominator disappear not only if we
put $\xi=0$, 
but also in a ``singular'' (in four dimensions)
gauge, $\xi=-4/(n-4)$. 
Having no $p_i^2$ in the denominator is convenient 
when one considers on-shell limits, i.e., when some of the
external momenta squared vanish;
otherwise, one needs to expand the scalar integrals
in the vanishing momenta squared (see Section~4).
Secondly, many terms vanish for $\xi=-2/(n-3)$, which could be
considered an $n$-dimensional generalization of the Fried--Yennie
gauge \cite{FriedYennie}.

In four dimensions, the only function which is ultraviolet-divergent
is the $A$ function. It should be renormalized by adding the
counterterm
\[
A^{(1,CT)} \!= \frac{{\overline{g}}^2}{(4\pi)^2} 
\left[ 
C_A \left( {\textstyle{2\over3}} \!+\! {\textstyle{3\over4}} \xi \right)
\!-\! {\textstyle{3\over4}} N_f T_R \right]
\left( \frac{1}{\varepsilon} \!+\! R \right)
\]
where $R$ is a constant related to a choice of renormalization
scheme, whereas 
${\overline{g}}^2 \equiv 
g^2 e^{-\gamma \varepsilon} (4\pi)^{\varepsilon}$
is the ``rescaled'' coupling constant. Such a re-definition of $g^2$ 
is usually performed in the ${\overline{\mbox{MS}}}$
renormalization scheme \cite{MSbar} which corresponds to the 
choice $R=0$
(cf. eq.~(15) of ref.~\cite{CG}).
In the symmetric limit, $p_1^2=p_2^2=p_3^2$, our results
(after renormalization) coincide with those from the paper \cite{CG}
(which had been also confirmed in \cite{PasTar}). 

\section{ON-SHELL LIMITS}

We first consider the case of {\it one} external momentum squared 
being zero, $p_3^2=0$.
Note that now we should consider the scalar functions $A$, $B$, $C$ 
and $F$  from (\ref{BC-ggg}) with permuted arguments as well. 
The result for the triangle integral (\ref{J(1,1,1)}) 
simplifies in this limit, 
\begin{eqnarray}
\label{J111_p3^2=0}
J(1,1,1) \bigg|_{p_3^2=0} 
= \mbox{i} \pi^{2-\varepsilon} \; \eta \; \varphi(p_1^2,p_2^2,0)  
\nonumber \\
= \mbox{i} \pi^{2-\varepsilon}
\; \eta \; 
\; \frac{1}{\varepsilon^2} \;
\frac{(-p_1^2)^{-\varepsilon} - (-p_2^2)^{-\varepsilon}}{p_1^2 - p_2^2} ,
\end{eqnarray}
where $\eta$ is defined by eq.~(\ref{eta}). Moreover,
in the framework of dimensional regularization \cite{dimreg},
$ J(1,1,0)|_{p_3^2=0} = 0$, 
while the results for $J(1,0,1)$ and $J(0,1,1)$ remain unchanged.
The $1/\varepsilon$ pole in (\ref{J111_p3^2=0})
corresponds to the infrared (on-shell) singularity which arises 
in the scalar integral in the limit $p_3^2=0$.

For the Feynman gauge, $\xi=0$ (and also for the singular gauge,
$\xi=-4/(n-4)$), it is enough to perform the above substitutions
to get the answer. In the case of arbitrary $\xi$, however, the 
situation is more tricky, due to the presence of $p_3^2$ in the
denominators of the scalar functions. 
Here, in order to get a correct answer, one needs the next term of
the expansion of
the integral $J(1,1,1)$ in $p_3^2$ (see in \cite{DOT}).

The infrared $1/\varepsilon$ singularities of the results for gluon 
and ghost contributions have been compared with the results given 
in \cite{BF}, eqs.~(24)--(25).
The functions $G^j$ defined in \cite{BF} 
can be represented as linear combinations of the scalar functions
(\ref{BC-ggg}), including those with permuted arguments. 
To get renormalized results, the counterterm $A^{(1,CT)}$ was 
added to all $A$ functions.  
In the ${\overline{\mbox{MS}}}$ scheme, the obtained results 
coincide with those presented in ref.~\cite{BF}, eq.~(25) 
(up to a minor misprint).

We next consider one external {\it momentum} being zero, $p_3=0$.
In this case, $p_1=-p_2\equiv p \;$ ($p_1^2=p_2^2=p^2$), 
and the proper limit of eq.~(\ref{J111_p3^2=0}) yields:
\begin{equation}
\left. \frac{}{} J(1,1,1) \right|_{p_3=0}
=  \mbox{i} \; \pi^{2-\varepsilon}\;
\; \eta 
\; \frac{1}{\varepsilon} \; (-p^2)^{-1-\varepsilon} .
\end{equation}
Actually, we get some powers of $(p_1^2-p_2^2)$ in the 
denominator from the ${\cal{K}}$'s, since 
${\cal{K}}=-\frac{1}{4} (p_1^2-p_2^2)^2$ in this limit. Therefore, we
should be careful taking the limit $p_2^2 \to p_1^2$
and expand the numerator up to higher powers in $(p_1^2-p_2^2)$.

In this limit, there are only three independent tensor structures left,
\begin{eqnarray}
\label{Gamma_p3=0}
\Gamma_{\mu_1 \mu_2 \mu_3}(p,-p,0)
\hspace{42mm}
\nonumber \\
= 2 \; g_{\mu_1 \mu_2} p_{\mu_3} \; 
\left[ A(p^2, p^2; 0) + p^2 \; C(p^2, p^2; 0) \right]
\nonumber \\
- \left( \! g_{\mu_1 \mu_3} p_{\mu_2} 
     \!+\! g_{\mu_2 \mu_3} p_{\mu_1} \! \right)
\left[ A(0, p^2; p^2) \!-\! B(0, p^2; p^2) \right]
\hspace*{-8mm}
\nonumber \\
- 2 p_{\mu_1} p_{\mu_2} p_{\mu_3} \; C(p^2, p^2; 0) \; .
\hspace{10mm}
\end{eqnarray}
The following relation holds for the zero-mo\-men\-tum case
(for arbitrary values of $n$ and $\xi$):
\[
A^{(1)}(p^2, p^2; 0) \!-\! A^{(1)}(0, p^2; p^2) 
\!+\! B^{(1)}(0, p^2; p^2) 
\!=\! 0.
\] 
Using this relation, 
we can reduce the number of tensor structures
in (\ref{Gamma_p3=0}) from three to two,
\begin{eqnarray}
\label{BL-decomp}
\Gamma_{\mu_1 \mu_2 \mu_3}^{(1)}(p, -p, 0)
\hspace{42mm} 
\nonumber \\
= \left( 2 g_{\mu_1 \mu_2} p_{\mu_3} \!-\! g_{\mu_1 \mu_3} p_{\mu_2}
         \!-\! g_{\mu_2 \mu_3} p_{\mu_1} \right) A^{(1)}(p^2, p^2; 0)
\hspace*{-7mm}
\nonumber \\
 + 2 p_{\mu_3} \left( p^2 g_{\mu_1 \mu_2} - p_{\mu_1} p_{\mu_2} \right)
              C^{(1)}(p^2, p^2; 0) ,
\hspace{5mm}
\end{eqnarray}
where the results for the scalar functions are given in \cite{DOT}.
Note that the first tensor structure on the r.h.s.\ of (\ref{BL-decomp})
coincides with 
(\ref{2eq:tree-level}).
Renormalizing our results in the $\overline{\mbox{MS}}$ scheme,
we find that they agree with those of \cite{BF}.  
We have also compared the renormalized version 
of eq.~(\ref{BL-decomp}) with the one-loop results
presented in ref.~\cite{BL}, eq.~(A10). 
We reproduce their result for $T_1$, but not for $T_2$.


Finally, we consider {\it two} external momenta squared being zero, 
$p_1^2=p_2^2=0$.
In this case, we find
\begin{eqnarray}
\label{J111_2}
J(1,1,1) \big|_{p_1^2=p_2^2=0}
= \mbox{i} \; \pi^{2-\varepsilon}\; \eta \; \varphi(0,0,p^2)
\nonumber \\
= - \mbox{i} \; \pi^{2-\varepsilon}\; \eta 
\; \frac{1}{\varepsilon^2} \; (-p^2)^{-1-\varepsilon} , \\
J(1,0,1) \big|_{p_1^2=p_2^2=0} 
= J(0,1,1) \big|_{p_1^2=p_2^2=0} = 0 ,
\end{eqnarray}
while the result for $J(1,1,0)$ remains unchanged.
Note that now, when two external lines are on-shell, the infrared
singularity in $J(1,1,1)$ is stronger and gives $1/\varepsilon^2$. 

Again, it is enough to make the above substitutions to get the
result in Feynman gauge ($\xi=0$), and in the singular gauge
($\xi=-4/(n-4)$), but the situation is more tricky for arbitrary $\xi$
since we have $p_1^2$ and $p_2^2$ in the denominators
of the scalar functions. To solve this problem, we need to
consider the expansion of $J(1,1,1)$ in $p_1^2$ and $p_2^2$.
Two independent ways were used to get the results for the scalar
functions in this limit: \\
(i) \ We take the expressions for one of the momenta squared
equal to zero (see above), and put the second momentum
squared equal to zero. In the corresponding expressions, all  
$p_i^2$ occurring in the denominators are always accompanied
by the corresponding $\kappa_i\equiv\kappa(p_i^2)$ in the 
numerator, which should be put equal to zero when $p_i^2$ vanishes. \\
(ii) \ First, we put $J(1,0,1)=J(0,1,1)=0$. Since the resulting expressions
did not have singularities for $p_1^2=p_2^2$, the next step was
to put $p_1^2=p_2^2\equiv p_0^2$. Then, the integral $J(1,1,1)$
was expanded in $p_0^2/p^2$ keeping the terms up to $(p_0^2/p^2)^2$.
Finally, the limit $p_0^2 \to 0$ was taken. 

The results obtained in these two ways coincide, 
the expressions obtained for the scalar functions (\ref{BC-ggg}) are
presented in \cite{DOT}. Infrared-divergent parts were successfully
compared with the expressions presented in \cite{BF}. 
The result for the three-gluon vertex in the Feynman gauge 
(for $p_1^2=p_2^2=0$) is available in Appendix~B of ref.~\cite{NPS}. 
It is expanded around $n=4$, and the divergent and finite (in $\varepsilon$)
parts are presented. In this limit, our expressions agree with
the results of \cite{NPS}. 

\section{CONCLUSIONS}

We have obtained 
results for the one-loop three-gluon vertex valid for arbitrary values
of the space-time dimension, $n$, and the covariant-gauge parameter,
$\xi$. We have considered the general off-shell case (arbitrary
$p_1^2, p_2^2, p_3^2$), as well as all on-shell
cases of interest. Moreover, having the results
in arbitrary dimension, it was possible to get all on-shell
expressions just by considering the corresponding limits of
the general (off-shell) results\footnote{The on-shell limits are
especially interesting for multijet calculations, see
e.g. \cite{BDK-review} and references therein.}. 
This would be impossible if one
started from the off-shell results expanded around $n=4$,
because in this case the infrared (on-shell) divergences would appear as
logarithms of vanishing momenta squared.

To calculate the vertex, we used the decomposition (\ref{BC-ggg}) 
and considered the six scalar functions, $A, B, C, S, F$ and $H$, 
which completely define the 
three-gluon vertex. One of these functions, namely the $S$ function,
was found to be identically zero at the one-loop order\footnote{
In the Feynman gauge and four dimensions, this results was obtained
in \cite{BC2}.}.
For special cases, we have 
successfully compared our results with those from the papers 
\cite{CG,BC2} (off-shell) and \cite{BF,NPS} (on-shell). 

Furthermore, in \cite{DOT} we have obtained general results
for the ghost-gluon vertex.
Employing these results, together with one-loop two-point 
functions, we have checked that the Ward--Slavnov--Taylor 
identity for the three-gluon vertex is 
(for arbitrary $n$ and $\xi$) satisfied by the expressions obtained. 
This is another non-trivial check on the longitudinal
part of the vertex (the $A$, $B$, $C$ and $S$ functions). 


{\bf Acknowledgements.} 
Research by A.~D.\ and P.~O.\ was supported by the Research Council of 
Norway. 
A.~D. is grateful to the organizers and to DESY-Zeuthen for supporting his
participation in the Rheinsberg Workshop.


\begin{thebibliography}{99}

\bibitem{QCD}
H.~Fritzsch, M.~Gell-Mann and H.~Leutwyler, 
         {Phys.\ Lett.} 47B (1973) 365; \\ 
D.J.~Gross and F.~Wilczek, {Phys.\ Rev.\ Lett.} 30 (1973) 1343; \\
H.D.~Politzer, {Phys.\ Rev.\ Lett.} 30 (1973) 1346; \\
S.~Weinberg, {Phys.\ Rev.\ Lett.} 31 (1973) 494.

\bibitem{CG} 
W.~Celmaster and R.J.~Gonsalves, {Phys.\ Rev.} D20 (1979) 1420.

\bibitem{BC2}
J.S.~Ball and T.-W.~Chiu, {Phys.\ Rev.} D22 (1980) 2550;
   Erratum: D23 (1981) 3085.

\bibitem{BF}
F.T.~Brandt and J.~Frenkel, {Phys.\ Rev.} D33 (1986) 464.

\bibitem{NPS}
M.A.~Nowak, M.~Prasza{\l}owicz and W.~S{\l}o\-mi{\'n}ski, 
     {Ann.\ Phys.} 166 (1986) 443.

\bibitem{DOT}
A.I.~Davydychev, P.~Osland and O.V.~Ta\-ra\-sov, 
Bergen Univ.\ preprint 1995-16, 
(hep-ph/9605348).

\bibitem{dimreg} 
G.~'t~Hooft and M.~Veltman, {Nucl.\ Phys.} B44 (1972) 189;\\
C.G.~Bollini and J.J.~Giambiagi, {Nuovo Cim.} 12B (1972) 20.

\bibitem{ibp}
F.V.~Tkachov, {Phys.\ Lett.}  B100 (1981) 65;\\
K.G.~Chetyrkin and F.V.~Tkachov, {Nucl.\ Phys.} B192 (1981) 159.

\bibitem{JPA}
A.I.~Davydychev, {J. Phys.} A25 (1992) 5587.

\bibitem{UD3}
N.I.~Ussyukina and A.I.~Davydy\-chev, {Phys.\ Lett.} B332 (1994) 159;
B348 (1995) 503.

\bibitem{DT2}
A.I.~Davydychev and J.B.~Tausk, Mainz/Ber\-gen Univ.\ preprint MZ-TH--95-14,
1995-07 (hep-ph/9504431), to appear in {Phys.\ Rev.}~D.

\bibitem{walnut}
R.~Scharf, Diploma Thesis, W\"urzburg, 1991; \\
C.~Ford, I.~Jack and D.R.T.~Jones, {Nucl.\ Phys.} B387 (1992) 373; \\
A.I.~Davydychev and J.B.~Tausk, {Nucl.\ Phys.} B397 (1993) 123.

\bibitem{'tHV-79}
G.~'t~Hooft and M.~Veltman, {Nucl.\ Phys.} B153 (1979) 365.

\bibitem{tensor}
L.M.~Brown and R.P.~Feynman, {Phys.\ Rev.} 85 (1952) 231;\\
G.~Passarino and M.~Veltman, {Nucl.\ Phys.} B160 (1979) 151.

\bibitem{reduce}
A.C.~Hearn,  {\em REDUCE User's Manual} (version 3.5),
RAND publication CP78  (Santa Monica, 1993).

\bibitem{FriedYennie} H.M. Fried and D.R. Yennie,
{Phys.\ Rev.\ (Ser.~2)}, 112 (1958) 1391.

\bibitem{MSbar} W.A.~Bardeen, A.J.~Buras, D.W.~Duke and T.~Muta,
{Phys.\ Rev.} D18 (1978) 3998.

\bibitem{PasTar} P.~Pascual and R.~Tarrach, {Nucl.\ Phys.}
B174 (1980) 123.

\bibitem{BL} E.~Braaten and J.P.~Leveille, {Phys.\ Rev.}
D24 (1981) 1369.

\bibitem{BDK-review} Z.~Bern, L.~Dixon and D.A.~Kosower,
Pre\-print SLAC-PUB--7111 (hep-ph/9602280), to appear in
Ann.\ Rev.\ Nucl.\ Part.\ Sci.

\end{thebibliography}
\end{document}